\documentclass[conference]{IEEEtran}
\IEEEoverridecommandlockouts
\usepackage{graphicx}
\usepackage{cite}
\usepackage{tabularx}
\usepackage{enumitem}
\usepackage{booktabs}
\usepackage{multirow}
\usepackage{xspace}
\usepackage{subcaption}
\usepackage{rotating}
\usepackage{framed}
\usepackage{url}
\usepackage{amsmath,amssymb,amsfonts}
\usepackage{algorithmic}
\usepackage{textcomp}
\usepackage{xcolor}
\usepackage{threeparttable}
\usepackage{flushend}
\usepackage{balance}
\usepackage{caption}
\usepackage{subcaption}

\def\BibTeX{{\rm B\kern-.05em{\sc i\kern-.025em b}\kern-.08em
    T\kern-.1667em\lower.7ex\hbox{E}\kern-.125emX}}

\graphicspath{{figures/}}

\newcommand{\hfh}{\textsc{Hugging Face Hub}\xspace}
\newcommand{\hfc}{\textsc{HFCommunity}\xspace}
\newcommand{\gh}{\textsc{GitHub}\xspace}
\newcommand{\github}{\textsc{GitHub}\xspace}
\newcommand{\gitlab}{\textsc{GitLab}\xspace}

\newcommand{\mycolor}{black}

\begin{document}

\title{On the Suitability of Hugging Face Hub for Empirical Studies} 

\author{\IEEEauthorblockN{Adem Ait}
\IEEEauthorblockA{IN3 -- UOC \\
Barcelona, Spain \\
aait\_mimoune@uoc.edu}
\and
\IEEEauthorblockN{Javier Luis C\'anovas Izquierdo}
\IEEEauthorblockA{IN3 -- UOC \\
Barcelona, Spain \\
jcanovasi@uoc.edu}
\and
\IEEEauthorblockN{Jordi Cabot}
\IEEEauthorblockA{IN3 -- UOC, ICREA \\
Barcelona, Spain \\
jordi.cabot@icrea.cat}
}

\maketitle

\begin{abstract}
\noindent\textsc{Background.}
The development of empirical studies in software engineering mainly relies on the data available on code hosting platforms, being \github the most representative.
Nevertheless, in the last years, the emergence of Machine Learning (ML) has led to the development of platforms specifically designed for developing ML-based projects, being \hfh (HFH) the most popular one.
With over 250k repositories, and growing fast, HFH is becoming a promising ecosystem of ML artifacts and therefore a potential source of data for empirical studies.
However, so far there have been no studies evaluating the potential of HFH for such studies.

\noindent\textsc{Objective.}
In this proposal for a registered report, we aim at performing an exploratory study of the current state of HFH in order to investigate its suitability to be used as a source platform for empirical studies.

\noindent\textsc{Method.}
We conduct a qualitative and quantitative analysis of HFH for empirical studies.
The former will be performed by comparing the features of HFH with those of other code hosting platforms, such as \gh and \gitlab.
The latter will be performed by analyzing the data available in HFH.
\end{abstract}


\section{Introduction}
\label{sec:introduction}
The development of empirical studies in Open-Source Software (OSS) requires large amounts of data regarding software development events and developer actions, which are typically collected from code hosting platforms.
Code hosting platforms are built on top of a version control system, such as Git, and provide collaboration tools such as issue trackers, discussions, and wikis; as well as social features such as the possibility to watch, follow and like other users and projects.
Among them, \gh has emerged as the largest code hosting site in the world, with more than 80 million users and 200 million repositories.

The emergence of Machine Learning (ML) has led to the development of platforms specifically designed for developing ML-based projects, being \hfh (HFH) one of the most popular ones.
HFH is a place where developers can publish and share their ML-based projects, as well as reuse datasets, pre-trained models and other ML artifacts.
As of April 2023, the platform hosts more than 250k repositories, and this number is growing fast.

In the last months, HFH has been evolving and incorporating features which are typically found in \gh, such the ability to create discussions or submit pull requests enabling more complex interactions and development workflows.
This evolution, its growing popularity and the ML-specific features make HFH a promising source of data for empirical studies.
Although the usage of HFH in empirical studies is promising, the current status of the platform may involve relevant perils.

In this paper, we propose a registered report to study the current state of HFH and its suitability to be used as a source for empirical studies.
We understand as suitability the amount and adequacy of the features to enable software development practices and the sufficient quantity of data to enable the conduction of empirical studies about such practices.
To this aim, we propose an execution plan where we analyze the set of features provided by HFH and then study the availability and quality of the data available in HFH.
Later, in the analysis plan, we discuss the results to evaluate their impact in different scenarios commonly found in empirical studies

The rest of the paper is structured as follows.
{\color{\mycolor}Sections~\ref{sec:background} and~\ref{sec:relatedwork} provide the background and related work, respectively.}
Section~\ref{sec:rqs} presents the research questions.
Section~\ref{sec:executionPlan} describes the execution plan, while Section~\ref{sec:variables} presents the variables identified in our report.
Sections~\ref{sec:analysisPlan} and~\ref{sec:threats} describe the analysis plan and the threats to validity, respectively.
Finally, Section~\ref{sec:conclusion} concludes the paper.

\section{Background}
\label{sec:background}
Hugging Face, the company behind HFH, is an AI company originally known for its Natural Language Processing (NLP) model called Hierarchical Multi-Task Learning (HMTL)~\cite{DBLP:conf/aaai/SanhWR19} or for the Transformers library~\cite{DBLP:conf/emnlp/WolfDSCDMCRLFDS20}, which provides APIs and tools to easily download and train state-of-the-art pretrained models.
Nevertheless, it became a household name thanks to the creation of HFH, its ML-based hosting platform, with the goal of building the largest open-source collection of ML artifacts to advance and democratize the access to ML for everyone.

HFH is a Git-based {\color{\mycolor}online} code hosting platform aimed at providing a hosting site for all kinds of ML artifacts, namely: 
(1) models, pretrained models that can be used with the Transformers library; 
(2) datasets, which can be used to train ML models; and 
(3) spaces, demo apps to showcase ML models.

The storage for these artifacts relies on Git repositories, where each repository is presented on the HFH website via three tabs, namely: card, files and community.
The repository card is the front face of the repository, and it is different for each repository type.
For model repositories, the card display the content of the \texttt{README} file, the downloads by month, the repository dependencies (datasets used for training, and spaces displaying the model) and there is an Inference API\footnote{https://huggingface.co/docs/api-inference/index} interface to test and evaluate the model.
For dataset repositories, it shows the repository dependencies, the \texttt{README} file and its downloads, along with a preview of the data.
The card for space repositories is the most different and changes from one space to another, as it is designed to provide a demo of an ML model.
Next to the repository card, the file tab displays the repository files and their commit history while the community tab hosts the discussions and pull requests threads arisen during the development of the repository.

\begin{figure}
    \centering
    \begin{subfigure}{\columnwidth}
        \centering
        \includegraphics[width=\columnwidth]{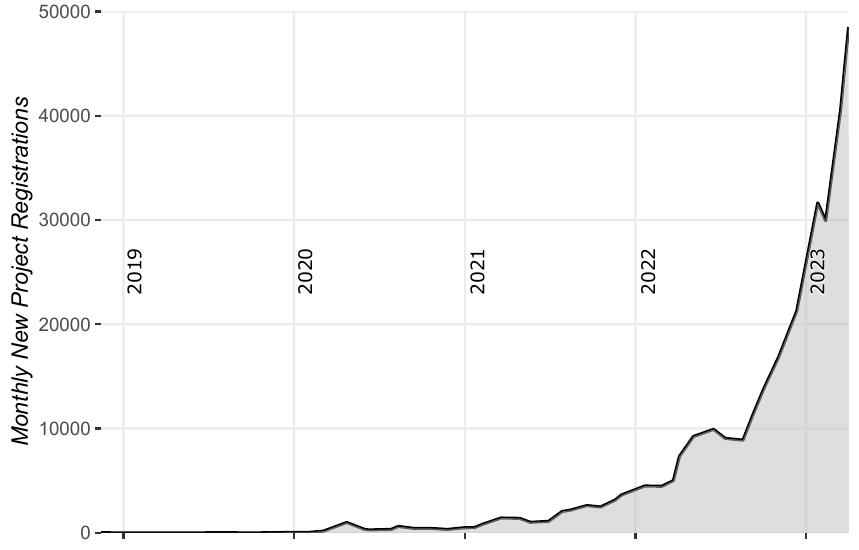}
        \caption{}
        \label{fig:numProjectsComparison:numprojectsHF}
    \end{subfigure}
    \begin{subfigure}{\columnwidth}
        \centering
        \includegraphics[width=\columnwidth]{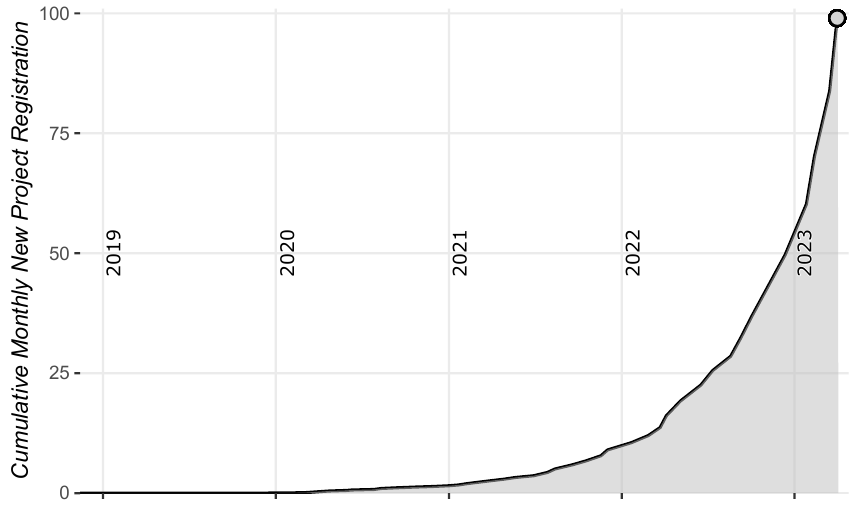}
        \caption{}
        \label{fig:numProjectsComparison:cumulativeHF}
    \end{subfigure}
    \begin{subfigure}{\columnwidth}
        \centering
        \includegraphics[width=\columnwidth]{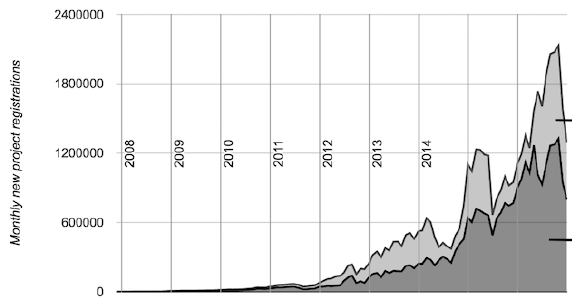}
        \caption{}
        \label{fig:numProjectsComparison:numprojectsGH}
    \end{subfigure} 
    \begin{subfigure}{\columnwidth}
        \centering
        \includegraphics[width=\columnwidth]{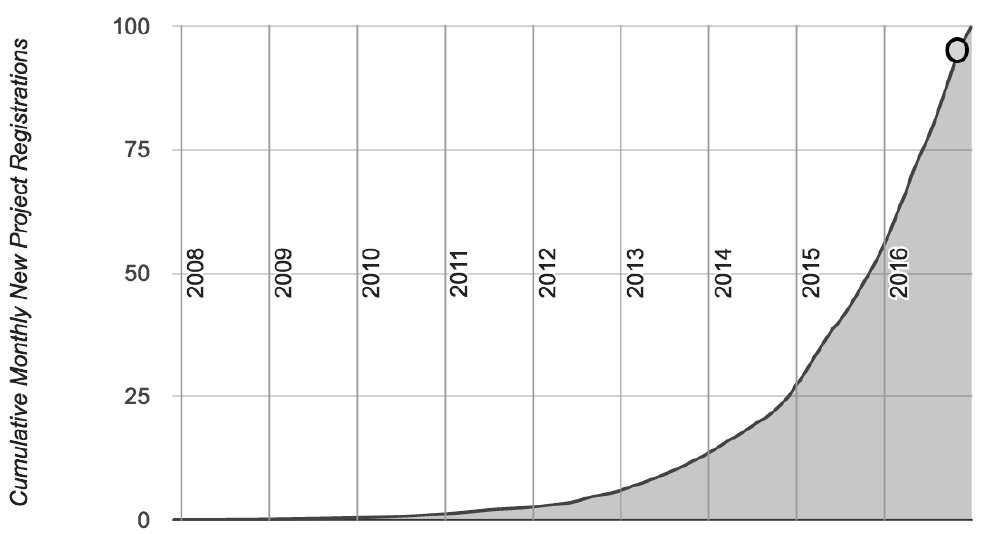}
        \caption{}
        \label{fig:numProjectsComparison:cumulativeGH}
    \end{subfigure}
    \caption{(a) Monthly and (b) cumulative new project registrations in HFH, 2018-2023. (a) Monthly and (b) cumulative new project registrations in \gh, 2008-2016.}
    \label{fig:numProjectsComparison}
\end{figure}

Since its creation, HFH has been rapidly evolving and incorporating new features.
For instance, the last tab regarding discussions was just released on May 2022.
{\color{\mycolor} 
To illustrate the growing evolution of the platform, Figures~\ref{fig:numProjectsComparison:numprojectsHF} and~\ref{fig:numProjectsComparison:cumulativeHF} illustrate the natural and cumulative growth of new project registrations by month in HFH, respectively.
To study the growth of the platform, we relied on the Diffusion of Innovation (DoI) theory~\cite{DBLP:books/daglib/0012785}, which helps to explain how a product gains or loses momentum in a system.
In Figure~\ref{fig:numProjectsComparison:cumulativeHF} the point indicates the month with the maximum growth.
As can be seen, the point is located in the last month of registered activity, which indicates that no momentum lost is detected, and therefore the platform is still growing.
For the sake of comparison, Figures~\ref{fig:numProjectsComparison:numprojectsGH} and~\ref{fig:numProjectsComparison:cumulativeGH} show the same growth for \gh, which was reported by Squire~\cite{DBLP:conf/wikis/Squire17}.
Note that HFH follows a growth similar to \textsc{GitHub}.
In the fifth year, the number of new projects registered in the platform was roughly the same.
Even though HFH is showing such a growing behavior, to the best of our knowledge, the number of research papers targeting empirical studies based on the platform is still very scarce.
}

\section{State of the Art}
\label{sec:relatedwork}
Many projects developed in code hosting platforms are public, thus allowing anyone to explore their activity, which include access to commits, issues, pull requests and comments, among others.
This large amount of public data has enabled researchers to easily collect and analyze such data.
As a result, many empirical studies have been conducted in the last years, in particular, mostly relying on the \gh platform~\cite{DBLP:conf/msr/DemeyerMWL13,DBLP:journals/access/CosentinoIC17,MSRSampling}.
 
However, the potential perils of empirical studies on this public software data are also relevant~\cite{DBLP:conf/msr/KalliamvakouGBSGD14,DBLP:conf/msr/HowisonC04,DBLP:journals/ese/KalliamvakouGBS16,DBLP:journals/ese/FlintC022}.
Perils could involve the quality of the project's data, the scarce use of the platform's features or the purpose of the project, etc.
This situation may affect to the quality of empirical studies, but also may raise concerns about the replicability of the results~\cite{DBLP:conf/msr/Robles10}.

In the last years, the number of ML-based projects has been increasing and motivated the SE community to understand the differences to traditional software development~\cite{DBLP:conf/msr/Gonzalez0N20}.
Also, platforms such as HFH have appeared to facilitate the development of ML-based projects and host their projects, models and data. Therefore, HFH could be regarded as a key platform for new studies covering the new breed of ML-based projects.
However, as any other code hosting platform, empirical studies on HFH could also bring its own set of perils and promises.
To the best of our knowledge, no study has been conducted to evaluate the quality and quantity of the data available in HFH, and the potential of the platform to be used in empirical studies. 
Starting this discussion is the purpose of this report.

\section{Research Questions}
\label{sec:rqs}
The {\color{\mycolor}goal} of our registered report is to assess {\color{\mycolor}the current state} of HFH and analyze its adequacy to be used in empirical studies.
In particular, we plan to address the following research questions: 

\vspace{0.25em} 
\noindent\emph{\textbf{RQ1. What features does HFH provide as a code hosting platform {\color{\mycolor}to enable empirical studies}?  }}
We plan to comprehend the key features that characterize HFH both for individual projects (i.e. features oriented towards end-users planning to use HFH for their software development projects) and at the platform level (i.e. to facilitate the retrieval and analysis of global HFH usage information). 
This analysis will allow characterizing the platform and identifying potential use cases for empirical studies.
Thus, we subdivide RQ1 further into:

\vspace{0.25em}
\textit{RQ1.1 What features HFH offers to facilitate the collaborative development of ML-oriented projects?}
This research question performs an exploratory study of the features offered by HFH to projects hosted in the platform. 
In this RQ, we focus on the features serving project development tasks, such as pull requests for managing code contributions or issue trackers for notifying bugs or requests.
To this aim, we plan to study current code hosting platforms to define a feature framework to be used as a reference framework to analyze the platform. 

\vspace{0.25em}
\textit{RQ1.2 What features HFH offers at the platform level to facilitate access to the hosted projects' data?}
In this research question we examine the features provided by HFH aimed at retrieving its internal data, derived from the activity of projects hosted in it.
Indeed, note that these features are not necessarily aimed at developing software projects in the platform (as it is the case of the features studied in RQ1.1) but at enabling the data collection from them.
We plan to include features to cover the platform infrastructure offered by HFH to access the data, such as APIs, and whether there are other solutions built by the community, such as datasets.
Furthermore, we are interested on identifying whether such infrastructure enables to collect information from each of the HFH features identified in RQ1.1. 
We believe the availability and easy access to the data in a code hosting platform is a relevant factor for researchers when selecting platforms for their empirical studies.

\vspace{0.5em}
\noindent\emph{\textbf{RQ2. How is HFH currently being exploited?}}
We are interested in studying how HFH is so far being used at platform and project levels.
In each level, we will analyze the data within two perspectives: volume and diversity.
To measure the volume we will define quantitative variables, such as the number of repositories and users at platform level; or the number of files, contributors and commits at project level.
On the other hand, to measure diversity we will define categorical variables, such as the programming languages used in the repositories or the type of contributions (i.e., issues or discussions) in the projects.
Note that while RQ1 focuses on the features provided by the platform, RQ2 analyzes its current usage, thus allowing to better understand the platform dynamics.
We subdivide RQ2 further into:

\vspace{0.25em}
\textit{RQ2.1 What is the current state of the platform data in HFH?}
In this research question we explore 
{\color{\mycolor} how HFH is used as a whole}.
Some examples of variables to be used in this research question are the number of repositories and the level of dependency between them as an example of volume and diversity.

\vspace{0.25em}
\textit{RQ2.2 What is the current state of the project data in HFH?}
In this research question we explore {\color{\mycolor} the usage of HFH at project level}. 
Thus, instead of the platform, the repository becomes the unit of study. 
The goal is to characterize the average (or averages if we detect different typologies) project on HFH via the analysis of their number of files and commits, number of users, its temporal evolution, etc.\looseness-1

{\color{\mycolor} To identify the goal, research questions and metrics we followed an approach similar to GQM~\cite{DBLP:books/daglib/0029933} methodology.
Section~\ref{sec:variables} addresses the metrics designed to answer each research question.}

\section{Execution Plan}
\label{sec:executionPlan}
We devise an execution plan to obtain the evidences required to answer our research questions.
The plan has the following steps that will be executed sequentially:

\begin{enumerate}
    \item Analyze features from current code hosting platforms. 
    \item Construct a feature framework from the features identified in the previous step.
    \item Characterize HFH with the feature framework.
    \item Analyze the options offered by HFH to retrieve the data.
    \item Conduct the data extraction process.
    \item Perform the data analysis of the HFH data.
\end{enumerate}

The execution plan detail is as follows.
\textcolor{\mycolor}{The first step is to analyze the features from current code hosting platforms.}
\textcolor{\mycolor}{Based on this analysis, we will categorize the identified features and use them as the basis for the comparison dimensions of a feature framework.}
\textcolor{\mycolor}{We will then instantiate HFH using this framework to compare HFH offering with those of other platforms}. These first three steps will address RQ1.1.

The next step is to analyze the exposure mechanisms offered by HFH to access the hosted dta.
The intention is to determine whether the data originated from projects using HFH features can be queried and, if so, {\color{\mycolor}the easiness of data retrieval (cf. Section~\ref{sec:analysisPlan}).} 
This will answer RQ1.2. 

Once addressed RQ1, the next two steps will target RQ2. 
The way to target RQ2 will depend on the results of the analysis of the RQ1.2. 
If the situation has not recently changed, it is likely that the best starting point for RQ2 is \hfc~\cite{hfc}, a curated dataset that periodically publishes up-to-date HFH data. 
Regardless of the source, we will complete the study with the data extraction and subsequent computation of several relevant metrics to understand the volume and diversity of HFH data, according to the perspectives mentioned in Section~\ref{sec:rqs}.

\section{{\color{\mycolor}Addressing the Research Questions}}
\label{sec:variables}
To address our research question we plan to conduct a qualitative and quantitative analysis of HFH.
The former will address RQ1, as it will focus on identifying the features of HFH and the options available to retrieve HFH data. 
The latter will address RQ2, and will allow us to analyze the data available in HFH via the reported data retrieval solution.
In the following we present the variables involved in each analysis.

\subsection{Qualitative analysis}
\label{sec:variablesQualitative}
During the qualitative analysis we will build a feature framework aimed at identifying which characteristics define a code hosting platform.
Features will include both characteristics offered to develop software projects and options provided to retrieve data from the platform.
{\color{\mycolor}The framework will be built by analyzing different code hosting platforms and identifying the features offered by each platform.}
{\color{\mycolor}We plan to revise a number of platforms, leveraging on existing literature (e.g.,~\cite{DBLP:journals/jsw/AlamerA17,DBLP:conf/wikis/Squire17}), but in the context of his paper we focus on \gh and \gitlab due to their relevance and wide user base.}
This will result in a superset of features \textcolor{\mycolor}{which will be backed by the literature from relevant venues} to underline the importance of some specific features in empirical studies.
{\color{\mycolor} Furthermore, we will validate the resulting set of features with the platform communities by conducting semi-structured interviews with relevant actors of each analyzed platform.} 
Note that the feature framework will help characterize HFH and, potentially, any code hosting platform by the features they offer.

Table~\ref{tab:featureFramework} shows a preliminar version of the feature framework, which may evolve during the realization of the registered report.
Features are identified as qualitative variables, and grouped according to topics.
The first five topics, namely: coding, social, user management, project management and project add-ons will address RQ1.1; while the last topic (i.e., platform) will address R1.2.
In the following, we describe each identified topic and motivate them in the context of empirical studies. 

\vspace{0.25em}
\noindent \textbf{Coding.} 
This topic includes variables addressing typical developers' needs to perform coding tasks, namely: usage of a version control software, and support for forks and pull requests, among others.
This would be the topic most related to the development process of contributors and has allowed the execution of empirical analysis of forking (e.g., \cite{DBLP:conf/icse/ZhouVK20,DBLP:conf/icis/NegoitaVSL19}), pull requests (e.g., \cite{DBLP:conf/icse/Subramanian20,DBLP:journals/ese/MezouarZZ19,DBLP:conf/apsec/LiRLZJ18}) or branching (e.g., \cite{DBLP:conf/qrs/ZouZ0H019})

\vspace{0.25em}
\noindent \textbf{Social.} 
This topic includes variables identifying user interaction and communication during development such as the creation of issues or Q\&A threads, and the ability to follow and like projects.
Addressing this kind of features has enabled studies on how users participate in discussions (e.g., \cite{DBLP:journals/ese/HataNBKT22}) or how issues are labelled during the development process (e.g., \cite{DBLP:journals/access/KimL21e,DBLP:conf/apsec/LiRLZJ18}). 

\vspace{0.25em}
\noindent \textbf{User management.} 
This topic is related to the ability of creating and managing groups of users with the purpose of sharing projects between multiple users.
Inside groups, a hierarchy structure can appear, thus defining roles inside the groups of users or inside a specific repository.
It can also enhance studies on how projects organize themselves~\cite{DBLP:conf/icse/VasilescuFS15,DBLP:conf/iisa/ChatziasimidisS15}.
Closely related to OSS development, these features may also help to conduct studies of OSS development roles inside repositories~\cite{DBLP:journals/ese/IzquierdoC22}. 

\vspace{0.25em}
\noindent \textbf{Project management.} 
The projects, or repositories, are the root element in social code hosting platforms.
This topic includes the study of support for project assets such as documentation or management tools such as milestones, labelling, or stream analytics.
Furthermore, we could also consider the chance of having different repository types, and relationships between them.
This kind of features have been used in empirical studies addressing survivability of projects~\cite{DBLP:conf/msr/AitIC22}, the use of milestones~\cite{DBLP:journals/smr/ZhangWWHW20} or checking security concerns~\cite{DBLP:conf/scored/BenedettiVM22}.

\vspace{0.25em}
\noindent \textbf{Project Add-ons.} 
Social code hosting platforms usually allow projects to integrate with apps, available via a marketplace; and communicate with external services via webhooks (e.g., \gh Actions).
This topic covers these features, which have enabled empirical studies on \gh \textsc{Actions}~\cite{DBLP:conf/icsm/DecanMMG22} or marketplace~\cite{DBLP:conf/sast/SouzaC0GB21}.

\vspace{0.25em}
\noindent \textbf{Platform.} 
This topic covers those auxiliary and technical tools to enable the collection of data for empirical studies. 
Thus, they aim at facilitating the use of the platform, including the existence of a platform API, an integrated CLI to access the platform or the indexation of the platform content, such a search mechanism.
Furthermore, we consider the existence of external datasets gathering data from the platform.

\begin{table*}[t]
\centering
\begin{threeparttable}
\caption{Qualitative analysis variables used in RQ1.}
\label{tab:featureFramework}
\begin{tabularx}{\textwidth}{lllX}
\textsc{RQ}             & \textsc{Topic}                      & \textsc{Feature}           & \textsc{Description}                                                               \\ 
\toprule
\multirow{28}{*}{RQ1.1} & \multirow{11}{*}{Coding}            & CVS                        & Control version system (e.g., Git, Mercurial, Subversion, etc.)                    \\ 
                        &                                     & Forking                    & Creation of a copy of other projects                                               \\ 
                        &                                     & Pull Request               & Submission of contributions to other projects                                      \\ 
                        &                                     & Code Review                & Discussion about changes in project files                                          \\ 
                        &                                     & Release                    & Identification and management of project releases                                  \\ 
                        &                                     & Packages                   & Management and release of \github packages                                         \\ 
                        &                                     & Snippets                   & Upload of fragments of code to share (e.g., \github Gist)                          \\ 
                        &                                     & Branches                   & Navigation and management of CVS branches                                          \\ 
                        &                                     & External integrations      & Integration with external services such as Campfire, Jira or Slack                 \\ 
                        &                                     & Collaborative/Cloud coding & Online development of project files (e.g., \github Codespaces, HTH resources)      \\ 
                        &                                     & Website publishing         & Support for serving HTML pages (e.g., \github Pages)                               \\ 
\cmidrule{2-4}
                        & \multirow{3}{*}{Social}             & Issues                     & Reporting of bugs and requests                                                     \\ 
                        &                                     & Q\&A                       & Discussions                                                                        \\ 
                        &                                     & Following                  & Support for stars and following platform users                                     \\ 
\cmidrule{2-4}  
                        & \multirow{2}{*}{User Management}    & Groups                     & Support for defining teams of users in projects                                    \\ 
                        &                                     & Roles                      & Roles inside the repository                                                        \\ 
\cmidrule{2-4}  
                        & \multirow{9}{*}{Project Management} & Milestone                  & Similar to Coding / Release                                                        \\ 
                        &                                     & Wiki                       & Wiki-based system for project's documentation                                      \\ 
                        &                                     & Work management            & Agile-like boards to organize tasks (e.g., \github projects; \gitlab To-do lists)  \\ 
                        &                                     & Stream analytics           & Project insights (e.g., \github analytics and repository insights)                 \\ 
                        &                                     & Tagging                    & Project's tag definition and management                                            \\ 
                        &                                     & Repository Type            & Classification of projects according to their purpose                              \\ 
                        &                                     & Project Relations          & Definition of link between projects (e.g., dependencies, etc.)                     \\ 
                        &                                     & Development workflows      & Continuous Integration and Development                                             \\ 
                        &                                     & Licensing                  & License identification for projects                                                \\ 
                        &                                     & Security                   & Access control to project's assets (e.g., visibility, code control, etc.)          \\ 
\cmidrule{2-4}  
                        & \multirow{2}{*}{Project Add-ons}    & Webhooks                   & Integration with external applications (e.g., \github Actions)                     \\ 
                        &                                     & Marketplace                & Catalogue of external integrations                                                 \\ 
\midrule        
\multirow{4}{*}{RQ1.2}  & \multirow{4}{*}{Platfom}            & Search                     & Search function for platform assets (e.g., repositories, files, users, etc.)       \\ 
                        &                                     & API                        & Support for accessing the platform programmatically                                \\ 
                        &                                     & Integrated CLI             & Tool to interact with the platform from the command line                           \\ 
                        &                                     & Datasets                   & Existing datasets to query the platform                                           \\ 

\bottomrule
\end{tabularx}
\end{threeparttable}
\end{table*}

\subsection{Quantitative analysis}
\label{sec:variablesQuantitative}

In the quantitative analysis we will examine the HFH data to provide an overview of the current usage of the platform.
We plan to study the usage of HFH at platform and project level.
The former will show the actual usage of the features identified in the previous research question and conclude on the level of exploitation of such features.
The latter will give an insight of how the development process is currently carried out in HFH, thus favoring the comprehension of why the users use this platform.
To address this analysis we leverage on the data provided by \hfc, an open-source tool that collects data from HFH and Git repositories, and stores it in a relational database to facilitate their analysis.
The database includes more than 250k repositories hosted in HFH.

We will define quantitative variables to analyze the platform and the repositories.
The variables are presented in Table~\ref{tab:variablesRQ2}.
The variables of the category \emph{Platform} will address the RQ2.1, while the variables of the category \emph{Project} will address RQ2.2.
The \emph{Platform} category is designed to characterize the HFH environment.
The HFH platform is specifically designed for ML-based artifacts, thus the platform aspects may differ, such as having repositories designed for each type of ML artifact.
Consequently, there can be variables specific for HFH.
For instance, we can study the nature of a repository type (i.e., a pre-trained model or a dataset) or the dependencies between the repositories, as one dataset can be used by multiple models.
We can measure the amount of repositories of each type or the proportion between datasets and models.

Regarding the \emph{Project} category, we will intend to give an insight on the status of the repositories.
As aforementioned, empirical analysis usually rely on a reduced subset of the repositories in a code hosting platform.
Therefore, we find appropriate to perform an analysis from a repository perspective and identify whether there is a way to select prolific repositories to perform empirical analysis as it is done on other code hosting platforms.
The variables in this category will intend to describe project specific characteristics, namely: 
\textcolor{\mycolor}{the purpose of the users to use this platform, by analyzing their activity and involvement in projects; the development process followed in a repository, via the information of activity and content; or the communication between contributors.
Furthermore, we propose demographic measurements to better describe a project, namely: the age, the project's artifact type, the number and type of dependent repositories, and the popularity, understood as the number of likes and downloads.}

\begin{table*}[t]
    \centering
    \caption{Quantitative analysis variables used in RQ2.}
    \label{tab:variablesRQ2}
    \begin{tabularx}{\textwidth}{llcX}
    \textsc{Category}                   & \textsc{Variable}                            & \textsc{Type}              & \textsc{Description} \\
    \toprule
    \multirow{4}{*}{Platform}           & Number of repositories                       & Q                          & Amount of projects in the platform \\                        
                                        & Diversity of repositories                    & C                          & \textcolor{\mycolor}{Distribution projects according to their category} in the platform \\
                                        & Number of users                              & Q                          & Amount of users in the platform \\   
                                        & Dependency of repositories                   & C                          & Communities \textcolor{\mycolor}{identified} by the dependency graph \\
    \midrule
    \multirow{8}{*}{Project}            & Activity                                     & Q                          & Activity events \textcolor{\mycolor}{(e.g., commits, files, users)} over time \\                     
                                        & Content                                      & C                          & \textcolor{\mycolor}{Distribution of file composition} of repositories \\
                                        & Involvement                                  & Q                          & Amount of users contributing in the repository \\
                                        & Interactions                                 & Q                          & Communication \textcolor{\mycolor}{(e.g., number of comments)} of users within the repositories \\
                                        & \textcolor{\mycolor}{Age}                    & \textcolor{\mycolor}{Q}    & \textcolor{\mycolor}{Time span of project's life} \\
                                        & \textcolor{\mycolor}{Artifact type}          & \textcolor{\mycolor}{C}    & \textcolor{\mycolor}{The ML task addressed by the proposed artifact (e.g., image classification, text generation, etc.)} \\ 
                                        & \textcolor{\mycolor}{Dependent Repositories} & \textcolor{\mycolor}{Q/C}  & \textcolor{\mycolor}{Amount and type of dependent repositories of the project} \\
                                        & \textcolor{\mycolor}{Popularity}             & \textcolor{\mycolor}{Q}    & \textcolor{\mycolor}{Amount of likes and downloads} \\
    \bottomrule
    \multicolumn{4}{l}{Q: Quantitative. C: Categorical.}
    \end{tabularx}
\end{table*}

\section{Analysis Plan}
\label{sec:analysisPlan}
Once we execute our study, we plan to analyze the results to explore the suitability of HFH for empirical studies. 
Our intention is to discuss how the current situation of HFH, assessed by the research questions, can enable empirical studies and help researchers to identify potential working lines in the platform.
To this aim, the following procedures will be followed.

\vspace{0.25em}
\noindent\textbf{RQ1.}
The outcome of RQ1 will characterize HFH in terms of features, and will allow us to analyze their importance in the context of empirical studies.
We plan to study the intersection of these features with the ones offered by other code hosting platforms, such as \gh and \gitlab, by analyzing existing literature.
The intersection can bring three scenarios:
(1) shared features, which would enable replicating in HFH empirical studies performed in other platforms;
(2) features not available in HFH, where we will study its impact and significance with regard to existing empirical studies;
and (3) exclusive features of HFH, where we can discuss opportunities where unique HFH features may open new applications for empirical studies.

RQ1.2 targets the retrieval data process from HFH, and during the analysis we plan to compare the means offered by the platform with those available in other code hosting platforms.
Being a relatively young platform, it is expected that HFH does not offer the same amount of prepackaged or curated datasets as other consolidated platforms, but we will study current options to perform this process.
{\color{\mycolor} We define the easiness of data retrieval with different aspects: 
(1) Usability of the mechanisms, which values the amount of existing solutions, such as APIs or datasets; 
(2) the accessibility of such solutions, whether they are open to everyone or if they require some sort of login, for instance; 
(3) the limitations, such as a token-based restriction;
 and (4) the recentness of the solutions, which focus on whether there is a temporal gap between the retrieved information and the platform data at the moment of extraction.}

{\color{\mycolor} The objective of the qualitative analysis for RQ1 is to derive conclusions from the data, keeping a clear chain of evidence~\cite{DBLP:books/daglib/0029933}. 
Our intention is to build and apply the feature framework, which will allow us to apply hypothesis generation techniques.
These techniques are intended to find hypothesis from the data, and the result of these techniques are the hypothesis as such.
Examples of hypotheses generating techniques are ``constant comparisons'' and ``cross-case analysis''~\cite{DBLP:journals/tse/Seaman99}.
The analysis will be conducted with an editing approach formalism, which means codes are defined based on findings of the researcher during the analysis.
}

\vspace{0.25em}
\noindent\textbf{RQ2.}
The results from RQ2 will be used to analyze HFH in terms of the identified categories, namely: platform (i.e., RQ2.1) and projects (i.e., RQ2.2).
The analysis of the former will help us to provide an overview of the behavior and evolution of the platform, while the latter may help us to characterize the typical HFH project.

Regarding the \emph{Platform} category, we expect to gather the absolute data and perform an analysis process to extract descriptive statistic indicators.
The descriptive statistic indicators will reflect the current status of HFH and the adequacy in terms of platform data.
Another aspect we will consider is the ecosystem characterization.
HFH may constitute ecosystems in a particular way, leveraging on its own features, as \gh does with its topics.

Once analyzed the platform, we will also analyze the project as a study unit.
Covered by the variables in the \emph{Project} category, we aim to visualize the exploitation of the repository features by its users.
Thus, the development process may differ from other code hosting platforms as this platform is designed for ML artifacts.
Also, the nature of the repository may show different signs and focus of activity.
For instance, instead of having continuous coding development, the repositories could also be used as a hosting site, being the community features (discussion threads) the source of activity of the repository.

{\color{\mycolor} With regard to the analysis of the quantitative variables for RQ2 we plan to perform the quantitative data interpretation proposed per Wholin et al.~\cite{DBLP:books/daglib/0029933}, which identify three steps, namely:
(1) descriptive statistics, where the data is characterized using descriptive statistics;
(2) dataset reduction, in which abnormal or false data points are excluded;
and (3) hypothesis testing, where the hypotheses of the experiment are evaluated statistically, at a given level of significance.

We plan to leverage on descriptive statistics to get a general view of the HFH data.
For instance, measures of central tendency, such as the mean, mode, and median, may help to understand the behavior of users in the platform as in the involvement of users in a project. 
As the target of this report is an exploratory study, instead of just excluding abnormal data points, we also plan to analyze them to bring some indicators (e.g., the number of empty repositories in the platform). 
After the exclusion of the abnormal data points, we plan to apply the research questions as presented previously. 
}

It is expected to find that HFH would not be suitable for longitudinal studies for long time projects or a massive number of projects, as we believe HFH is not mature enough to be treated as a large code hosting platform.
The conclusions of this research question will aid us to be able to sustent such discussions.

\section{Threats to Validity}
\label{sec:threats}
Our work is subjected a number of threats to validity, namely: 
(1) internal validity, which refers to the inferences we will make;
(2) external validity, which is related to the generalization of our findings;
(3) construction validity, which refers to the approaches we use to address the research questions;
and (4) conclusion validity, which is related to the interpretation of our results.

\subsection{Internal \& External Validity}

Regarding the internal validity, to address RQ1 we relied on our feature framework which may not cover all the features from code hosting platforms.
The dimensions of our framework will be gathered by an analysis of a set of platforms.
However, these platforms provide subsets of features according to their business objectives. 
Furthermore, the interpretation of the features and topics is subjected to the understanding of the authors.
To address RQ2 we will rely on the data provided by \hfc, which uses data from the HFH API and Git.
Git and HFH data may suffer from user clashing, as usernames in both platforms may not match, as reported by Ait et al.~\cite{hfc}.
\textcolor{\mycolor}{It is also important to note that the emerging behavior of HFH does not demonstrate consolidation and spread yet, which may limit the scope of our inferences.}

As for the external validity, the analysis done in RQ1 will rely on the current feature offer of HFH at the moment of performing our registered report, but it may change in the future.
On the other hand, the dataset used in RQ2 will be based on a set of HFH projects from \hfc, which releases monthly versions.
Thus, the results of the registered report should not be directly generalized without proper comparison and validation.
{\color{\mycolor} In particular, it is important to note that the results of this study have been obtained from a particular snapshot of HFH.}

\subsection{Construct \& Conclusion Validity}

The process to construct the feature framework will follow an iterative approach where each social code hosting platform is analyzed to identify the features.
In this approach, each platform is studied individually to identify its features, and then they are shared to identify a superset of features
As some features may be shared or be similar, the process repeats until no more new features are identified.
In the last step, the set of features are grouped according to a topic.
\textcolor{\mycolor}{The main process is performed by the first author, and the results are debated by the second and third authors.
Disagreements are discussed until a consensus is reached.
As mentioned in Section~\ref{sec:variablesQualitative}, the resulting set of features will be validated by conducting semi-structured interviews with relevant actors of each analyzed platform}

Conclusion validity affects to the results of the analysis plan, which relies on the results of the execution plan.
In this case, the conclusion validity is mainly threatened by biases of our interpretation of the results of the RQs.

\section{Conclusion}
\label{sec:conclusion}
In this registered report we presented our proposal for a study of the suitability of HFH for empirical studies.
The study comprises a feature-based framework comparison to characterize the HFH functionality together with an analysis of the mechanisms to retrieve information on how such features are used. 
This allows evaluating the suitability of HFH from a feature availability perspective.
Besides this feature-level study we propose to conduct a second one, more quantitative one, based on the study of volume and diversity of the data stored on the HFH.
We conduct this study both at the platform and project-levels, looking at the overall volume and richness of the data and on how the average project uses the platform.

Beyond a deeper understanding on how collaborative development of ML-related projects takes place on the HFH, the conclusion of this report is expected to be a discussion on whether HFH can be a suitable data source to perform empirical studies.
Also, given that empirical studies usually focus on a specific characteristic of code hosting platforms, as we mentioned in Section~\ref{sec:variablesQualitative}, 
beyond a boolean answer, the goal is to discuss what types of empirical studies could benefit from HFH data, either as a standalone data source or in combination with GitHub or other data sources.

\section*{Acknowledgements}

This work is part of the project TED2021-130331B-I00 funded by MCIN/AEI/10.13039/501100011033 and European Union NextGenerationEU/PRTR; and BESSER, funded by the Luxembourg National Research Fund (FNR) PEARL program, grant agreement 16544475.

\bibliographystyle{IEEEtran}
\bibliography{rr-main}         

\end{document}